\documentclass[apj]{emulateapj}

\slugcomment{Accepted for publication in the Astronomical Journal 01/16/15}

\shorttitle{Spectroscopic Abundances in NGC 6819}
\shortauthors{Lee-Brown, Anthony-Twarog, Deliyannis, Rich, Twarog}

\begin{document}

\title{Spectroscopic Abundances in \\
the Open Cluster, NGC 6819 
\thanks{WIYN Open Cluster study LXV}}

\author{Donald B. Lee-Brown and Barbara J. Anthony-Twarog\altaffilmark{1}}
\affil{Department of Physics and Astronomy, University of Kansas\\ Lawrence, KS 66045-7582, USA \\donald@ku.edu, bjat@ku.edu}

\email{WIYN Open Cluster Study LXV}

\author{Constantine P. Deliyannis\altaffilmark{1}}
\affil{Department of Astronomy, Indiana University\\ Bloomington, IN 47405-7105, USA}

\author{Evan Rich\altaffilmark{2}}
\affil{Department of Physics and Astronomy, University of Kansas \\Lawrence, KS 66045-7582, USA}

\and
\author{Bruce A. Twarog}
\affil{Department of Physics and Astronomy, University of Kansas \\Lawrence, KS 66045-7582, USA}

\altaffiltext{1}{Visiting Astronomer, Kitt Peak National Observatory, National Optical Astronomy Observatory, which is operated by the Association of Universities for Research in Astronomy (AURA) under cooperative agreement with the National Science Foundation.}
\altaffiltext{2}{Current affiliation: Department of Physics \& Astronomy, University of Oklahoma, Norman, OK 73019, USA}


\begin{abstract}
High-dispersion spectra of 333 stars in the open cluster NGC 6819, obtained using the HYDRA spectrograph on the WIYN 3.5m telescope, 
have been analyzed to determine the abundances of iron and other metals from lines in the 400 \AA\ region surrounding the Li 6708 \AA\ line. 
Our spectra, with signal-to-noise per pixel (SNR) ranging from 60 to 300, span the luminosity range from the tip of the red giant branch 
to a point two magnitudes below the top of the cluster turnoff.  We derive radial and rotational velocities for all stars, as well as [Fe/H] 
based on 17 iron lines, [Ca/H], [Si/H], and [Ni/H] in the 247 most probable, single members of the cluster. Input $T_{eff}$ estimates for model 
atmosphere analysis are provided by $(B-V)$ colors merged from several sources, with individual reddening corrections applied to each 
star relative to a cluster mean of $E(B-V)$ = 0.16. Extensive use is made of ROBOSPECT, an automatic equivalent width measurement 
program; its effectiveness on large spectroscopic samples is discussed. From the sample of likely single members, [Fe/H] = $-0.03 \pm 0.06$, 
where the error describes the median absolute deviation about the sample median value, leading to internal precision
for the cluster below 0.01 dex. The final uncertainty in the cluster abundance is therefore dominated by external systematics due to
the temperature scale, surface gravity, and microturbulent velocity, leading to [Fe/H] = $-0.02 \pm 0.02$ 
for a sub-sample restricted to main sequence and turnoff stars. 
This result is consistent with our 
recent intermediate-band photometric determination of a slightly subsolar abundance for this cluster.
[Ca/Fe], [Si/Fe], and [Ni/Fe] are determined to be 
solar within the uncertainties. NGC 6819 has an abundance distribution typical of solar metallicity thin disk stars in the solar neighborhood.

\end{abstract}



\section{Introduction}

NGC 6819 is an old (2.3 Gyr) open cluster whose fundamental properties have garnered increasing attention 
in recent years. As detailed in \citet{AT14}, (hereinafter Paper I), its age places it in a sparsely populated 
range of open cluster ages, making it an invaluable testbed for stellar models of stars just above the Sun's mass.
It was chosen as a key cluster in our 
program to map the evolution of Li among stars of varying mass as they evolve from the main sequence 
to the tip of the giant branch and beyond. Delineation of such a map requires reliable estimates of stellar 
temperature, luminosity, and metallicity, especially iron, and, indirectly, the cluster reddening, distance modulus and age.

Data on individual cluster stars and the surrounding field have expanded due to the cluster's location in the {\it Kepler} 
field, bringing the added potential for asteroseismic insight into the structure of individual stars in a variety of 
evolutionary states \citep{GI10}. Reliable radial velocities are available for a large sample of stars extending 
well below the cluster turnoff \citep{H09} (hereinafter H09) and improved proper-motion memberships by \citet{PL13}
(hereinafter PL) have supplanted the older data of \citet{SA72}. Comprehensive broad-band \citep{RV98, KA01, YA13} and 
intermediate-band (Paper I) photometric surveys allow precise determination of each star's position in the color-magnitude 
diagram (CMD), while definitively demonstrating (PL, Paper I) that the cluster suffers from variable foreground reddening
with a range of $\Delta$$E(B-V)$ = 0.06 mag, not surprising given the cluster's galactic coordinates 
$(l,b = 74\arcdeg, +8\arcdeg$) and a distance of 2.4 kpc from the Sun. Until recently, the weakest link in the discussion of the 
properties of the cluster and its stars has been the cluster metallicity. As detailed in Paper I, metallicity estimation 
from cluster members has been plagued by small number statistics often coupled with incorrect assumptions of 
uniform and/or anomalously high reddening, leading to a consensus view that NGC 6819 was metal-rich, with [Fe/H] $\sim$ +0.1. 
Precise intermediate-band photometry on the extended Stromgren system (Paper I) strongly contradicts this claim, 
producing [Fe/H]$ = -0.06 \pm 0.04$ from 278 single, unevolved F stars. 

Although a fuller discussion of the Lithium abundances for our large sample in NGC 6819 is in preparation \citep{d15}, one preliminary result - the
identification of a Li-rich giant star - was reported in \citet{AT13}.  We employ a combination of spectral synthesis and 
curve-of-growth analysis to convert measured equivalent widths for the Li line at 6707.8 \AA\ to Lithium abundances, A(Li).  
Our curve-of-growth analysis method depends on subtracting an estimated contribution by an iron line at 6707.4 \AA\ from the equivalent width, 
for which accurate temperature and iron abundance information is required.
Given the criticality of a precise estimate of the stellar Fe abundance for derivation of Li from spectroscopy, especially 
among the giants, it was decided that all measurable lines within the wavelength range of our HYDRA high dispersion 
spectra of the Li 6708 \AA\ region would be used to estimate the cluster metallicity for Fe and other elements. 
In addition to testing the photometric abundance, following the pattern laid out in our earlier investigations 
of the open clusters NGC 3680 \citep{AT09} and NGC 6253 \citep{AT10}, such data could reveal the effects of stellar 
evolution on surface abundances through comparison of the red giants to the turnoff stars, while 
pinpointing the place of the cluster within the global context of galactic chemical evolution. 

The layout of the paper is as follows. Sec. 2 details the spectroscopic observations and reduction of
images to uniformly wavelength-calibrated and continuum-fitted spectra. 
Sec. 3 explains the collation of radial-velocity, photometric, proper-motion, and 
reddening data used to define the crucial temperatures, surface gravities, and microturbulent velocities for single-star 
cluster members that, together with appropriate model atmospheres, translate the spectroscopic measures to abundances. 
Sec. 4 supplies a detailed examination of ROBOSPECT, an automated line-measurement program used for this study which 
makes internally consistent and rapid abundance estimation for hundreds of stars feasible. Sec. 5 presents the abundance 
results for individual stars and the cluster as a whole, delineating the impact of the parametric uncertainties on our 
final results, and places the data within the context of both stellar and galactic evolution. Sec. 6 summarizes our conclusions.

\section{Spectroscopic Data}

Spectroscopic data were obtained for target stars in NGC 6819 using the WIYN 3.5-meter telescope \footnote{The WIYN Observatory is a joint facility of the University of Wisconsin-Madison, Indiana University, Yale University, and the National Optical Astronomy Observatory.}
and HYDRA 
multi-object spectrograph over 13 nights from September and October 2010, June 2011 and February 2013. The positions
of the target stars within the cluster $(V, B-V)$ CMD are shown in Figure 1.
Six configurations were designed to position fibers on a total of 333 stars. The brightest configurations 
include stars near $V \sim 11$, while the fainter configurations reach to $V \sim 16.5$. Individual exposures 
ranged from 10 to 90 minutes, with accumulated totals of 2.5 to 4.5 hours for stars in the brightest configurations, 
10 to 13 hours for the configuration with stars of intermediate brightness, and over 14 hours total for the 
faintest configurations. Our spectra cover a wavelength range $\sim$ 400 \AA\ wide centered on 6650 \AA, with 
dispersion of 0.2 \AA\  per pixel. Examination of Thorium-Argon lamp spectra indicates that the line resolution 
comprises 2.5 pixels, yielding a spectral resolution over 13,000.  

The data were processed using standard reduction routines in IRAF\footnote{IRAF is distributed by the National 
Optical Astronomy Observatory, which is operated by the Association of Universities for Research in Astronomy, 
Inc., under cooperative agreement with the National Science Foundation.}. These routines included, in order of 
application, bias subtraction, division by the averaged flat field, dispersion correction through interpolation of 
the comparison spectra, throughput correction for individual fibers using daytime sky exposures in the same configuration, and continuum 
normalization. After flat field division and before the dispersion correction, the long-exposure program images 
were cleaned of cosmic rays using ``L. A. Cosmic''\footnote{http://www.astro.yale.edu/dokkum/lacosmic/, an IRAF 
script developed by P. van Dokkum (van Dokkum 2001); spectroscopic version.}
\citep{VD01}. Real-time sky subtraction was accomplished by using the dozens of fibers not assigned to stars and 
exposed to the sky for each integration.  Composite spectra for each of the six configurations were 
constructed by additive combination. 

The signal-to-noise ratio per pixel (SNR) may be estimated two ways: first, by direct inspection of the 
spectra within IRAF's SPLOT utility, using mean values and r.m.s. scatter from a relatively 
line-free region or, second, by construction from output files of the ROBOSPECT software suite. We found the ROBOSPECT 
values to be entirely consistent with hand-measured SNR values and quote the ROBOSPECT-derived values in Table 1, computed 
from the relatively line-free 6680-6694 \AA\ region. SNR estimates 
reflect the statistics characterizing the summed composite spectra.

\section{Stellar Properties}

\subsection{Radial Velocities and Proper Motions}
An initial spectroscopic sample of probable cluster members was taken from the valuable radial-velocity survey of NGC 6819 
by H09. All stars brighter than $V \sim 16.75$ with radial-velocity membership probabilities greater than 50\% were 
identified as candidates for the present study. Stars classed as double-lined spectroscopic binaries were eliminated; 
single-lined systems were retained since the existence of the companion would have minimal impact on line measurement. 
Stars were not eliminated based upon their position in the CMD to avoid biasing the sample against 
stars undergoing anomalous evolution. 

Individual stellar radial velocities were derived from each summed composite spectrum utilizing the Fourier-transform, 
cross-correlation facility FXCOR in IRAF. In this utility, program stars are compared to stellar templates of similar 
effective temperature ($T_{eff}$) over the wavelength range from 6575 \AA\ to 6790 \AA, as well as a narrower region 
in the vicinity of H$\alpha$ alone. Typical uncertainties in the individual radial velocities were estimated at 1.15 km/sec. 
The FXCOR utility also provides measures of the line widths within each spectrum, from which rotational velocities may be inferred.
Observations of a pair of radial-velocity standards were obtained during each run and processed using the same procedure 
applicable to the cluster. Comparison of the velocity zero-points using standard values from the General Catalog of Radial 
Velocities \citep{WI53} allowed transformation of the cluster data to the standard system, within the uncertainties of the measurements.

As a first check on our spectroscopic data, we can compare our measured radial velocities with those of H09 to
identify discrepant stars or long-term variables. Eliminating 29 stars classed as spectroscopic binary members of 
NGC 6819, the remaining 304 single stars have a mean radial velocity of 2.65 $\pm$ 1.36 (s.d.) km/sec. The same sample from H09 
has a mean velocity of 2.38 $\pm$ 0.99 (s.d.) km/sec. The dispersion in the residuals between these samples is 1.06 km/sec; 
the predicted dispersion from the quoted errors for each star is 1.19 km/sec. The residuals for only three stars (WOCS 1007, 2016 and 56018) fall 
more than three sigma from the mean; one of these, WOCS 2016, has unusually broad lines (presumably due to high rotational velocity) which 
leads to a larger than average uncertainty in the final radial-velocity estimate. By contrast, the sample of 29 
spectroscopic binaries has a mean cluster velocity of 3.45 km/sec and a dispersion of
9.8 km/sec; the same stars from H09 exhibit a dispersion of 9.78 km/sec. We conclude that all single star members as classified 
by H09 are confirmed as such by our data. Table 1 contains the derived mean radial and estimated rotational velocities for each star 
in our survey.

At the start of our program, the only NGC 6819 proper-motion study available was that of \citet{SA72}, which proved inadequate 
for reliable identification of probable members, particularly at the fainter limit of interest. Fortunately, the comprehensive
survey by PL covers the appropriate range in both area and depth, generating probabilities for all stars in our sample except one.
Of 332 stars selected via radial velocity, 59 have proper-motion membership probabilities below 50\%, 43 of which 
are in single digits. The PL values for individual stars may be found in Table 1. The different stellar categories are identified
in the CMD, 
Fig. 1.

\subsection{Effective Temperature, Surface Gravity, and Microturbulent Velocity}

In keeping with our approach to spectroscopic abundance determination for previous clusters in this program, our default
scheme for determining model atmosphere input temperatures is based upon photometric color, 
specifically $B-V$. Four sources of broad-band $BV$ data exist. As a first
step, we merged the CCD photometry presented by H09 as PHOT98 and PHOT03. From 923 stars brighter than $V$ = 16.7, 
excluding 5 stars with absolute residuals greater than 0.10 mag, the mean offsets,
in the sense (PHOT98 - PHOT03), in $V$ and $B-V$ are $-0.005 \pm 0.022$ and $+0.003 \pm 0.022$, respectively.
While the dispersion in both sets of residuals was satisfactory, further analysis revealed that a large fraction of the
scatter in $V$ is the result of a nonlinear radial gradient among the residuals, reaching $\Delta V = -0.05$ mag for stars 
near the core of the cluster but $+0.01$ for stars in the outer regions of the frame. \citet{MI14} have independently 
discovered the same effect in a comparison between the $VI$ photometry of \citet{YA13} and PHOT03, implying that 
the primary source of the trend in the current comparison must lie with the PHOT03 database. Additional evidence 
in support of this conclusion comes from a direct comparison between the $V$ magnitudes from \citet{YA13} and 
the intermediate-band data from Paper I where the residuals do not exhibit a radial dependence. More important from 
our perspective is the absence of a radial trend among the residuals in $B-V$ between the two samples, PHOT98 and PHOT03. 
Application of a small color term, $(B-V)_{03} = 0.986(B-V)_{98}+ 0.006$, reduces the scatter in the residuals 
to $\pm 0.021$ mag. The PHOT98 $B-V$ data were converted to the PHOT03 system and averaged for stars common to both samples.

Next, the composite H09 data were transformed to our adopted standard of \citet{RV98}. From 470 stars with $V$
brighter than 16.5 common to the two datasets, removing 10 stars with residuals in $B-V$ greater than 0.05 mag, 
the mean residual, in the sense (RV-H09), is +0.006 $\pm$ 0.014 mag. A weak color dependence among the residuals was found, 
$(B-V)_{RV} = 1.007(B-V)_{H09}$, and applied prior to the merger of the two databases, eliminating the small zero-point offset.

Finally, a transformation was derived between the $B-V$ indices of \citet{KA01} and the previously merged $B-V$ data 
on the \citet{RV98} system; comparisons of the $V$ mags of \citet{RV98} and \citet{KA01} have already been discussed in Paper I.

Eliminating 11 stars with absolute residuals greater than 0.1 mag and applying a small color term, 
$(B-V)_{RV} = 1.025(B-V)_{KA} - 0.018$, 846 stars brighter than $V = 16.5$ exhibit a mean residual in $B-V$ of $0.000 \pm 0.018$ mag.

All stars with absolute residuals greater than 0.05 mag were individually checked. If $B-V$ estimates
were available from more than two sources and one source was clearly the origin of the discrepancy, that value was dropped.
If only two sources of $B-V$ existed, an independent check on the predicted $B-V$ was attempted using the published $V-I$
data of \citet{YA13}. In cases where no resolution of the discrepancy was possible, the averaged value from all sources was 
retained. Averaged $B-V$ indices from the four transformed primary data sets are given in Table 1 along with the number of 
sources and each star's estimated individual reddening correction.

As noted earlier, PL identified and mapped variable reddening across the field of NGC 6819, a result confirmed in Paper I.  
We used the map of individual reddening values derived by PL to estimate by spatial interpolation the degree of reddening 
affecting each of the stars in our spectroscopic sample.  Individual reddening estimates, with a range of 
$\pm 0.033$ about our adopted mean value of $E(B-V)=0.16$, were applied to each star's $(B-V)$ color. 
A temperature for each star based on its dereddened $(B-V)$ color was then derived using two primary color-temperature calibrations. 

For dwarfs, we continue to use a calibration \citep{DE02} consistent with previous spectroscopic studies by this group, 
namely: 

$T_{eff} = 8575 - 5222.7 (B-V)_0 + 1380.92 (B-V)_0^2 + 701.7 (B-V)_0[{\rm [Fe/H]} - 0.15]$ K.

\noindent
For giants, the $T_{eff}$-color-[Fe/H] calibration of \citet{RA05} was used. We note that for 13 stars, it was possible to obtain temperatures from both color-temperature calibrations.  Temperatures derived from the \citet{DE02} calibration are $41 \pm 6$ K higher for these mostly subgiant stars.
 
Surface gravity estimates (log $g$) were obtained by direct comparison of $V$ magnitudes and $B-V$ colors 
for our sample of 333 stars to isochrones from the $Y^2$ compilation \citep{DE04}, constructed for a scaled solar 
composition with [Fe/H] = -0.06 and an age of 2.3 Gyr, essentially the same as the comparison presented 
in Paper I. The isochrone's predicted magnitudes and colors were adjusted to match the cluster's reddening, 
$E(B-V)=0.16$ and apparent distance modulus, $12.40$.  For stars that appear to be blue stragglers, surface gravities 
were estimated by comparing photometric information to a grid of younger isochrones of similar composition.  

We are fortunate to have independent confirmation of the log $g$ values for the giant branch stars, thanks to 
their status as {\it Kepler} objects of interest. We compared log $g$ values from our isochrone match to log $g$ values 
inferred using asteroseismology from {\it Kepler} data, as presented by \citet{BA11} for 21 giants. The seismological gravities are in remarkable
agreement with the isochrone-inferred gravities, differing by an 
insignificant amount, $-0.02 \pm 0.04$ where the listed error is the standard deviation.

Input estimates for the microturbulent velocity parameter were constructed using various prescriptions. For dwarfs 
within appropriate limits of $T_{eff}$ and log $g$, the formula of \citet{ED93} was used; a similar formulation 
by \citet{RA13} extends to slightly lower temperatures and gravities and was used for some stars.  For giants, 
a gravity-dependent formula, $V_t= 2.0 - 0.2$ log $g$, was used.  For some subgiants and candidate blue stragglers, 
no suitable formula for $V_t$ was found other than the purely gravity-dependent expression employed by the SDSS 
collaboration in their DR10 data release and discussion of the abundance analysis pipeline for APOGEE spectra \citep{AH14}.

\section{Spectroscopic Processing With ROBOSPECT}
With the expansion in spectroscopic samples from a few dozen stars in previous cluster work \citep{AT09, AT10} to a few
hundred in this and future analyses, manual measurement of equivalent widths (EW) for individual lines is now prohibitive.
To overcome this obstacle, the automated line-measuring program, ROBOSPECT \citep{WH13} (hereinafter WH), has been
utilized. For full details, the reader is referred to WH; a brief outline of the program operation and our procedure is provided here.
ROBOSPECT measures EW by first determining the continuum level and constructing a noise profile across the wavelength 
range of interest using an iterative process. With the initial continuum and noise levels established, spectral lines are tagged 
at locations specified by a user-supplied line list, and other potential lines are automatically identified based on significant 
deviations of the local spectrum from the continuum. For the automatic line identification, the significance threshold is 
user-specified and based on the current noise solution. After lines are identified though these two processes, they are 
subtracted from the spectrum and the continuum process is repeated and the noise level refined. Through multiple iterations, 
the best-fit continuum and individual line solutions are reached. For this study, each spectrum was individually corrected 
in ROBOSPECT for radial velocity and run through 25 iterations of continuum fitting and line estimation using a gaussian 
line profile with three-sigma automatic line identification and no least-squares line deblending. All other parameters for 
the program were set to default values.      

An issue of obvious concern with any automated procedure is the potential for unreliable EW due to low SNR spectra, line blending,
or inaccurate radial-velocity correction. With interactive measurement, such issues can be flagged and corrected or eliminated 
but, with ROBOSPECT, flawed EW measures propagate to followup steps in the abundance determination process, potentially 
distorting the final results. It is expected that the uncertainty in EW will increase with decreasing SNR, potentially reaching a
level where the subsequent uncertainty in a star's calculated abundance renders it useless. Similarly, we can expect the EW 
uncertainty to increase as individual lines suffer increased blending from neighboring lines. Finally, a significant error in
radial velocity will cause ROBOSPECT to misidentify the wavelength of line centers. To ensure that our abundances are minimally
affected by these issues, a variety of tests were conducted to determine if our line list and ROBOSPECT input parameters 
lead to robust results over the range in SNR and radial-velocity uncertainties spanned by our spectra. Since ROBOSPECT will be
used in future analyses with similar input parameters, the level of detail supplied is somewhat greater than usual.

The line list used during this study was constructed by visually identifying relatively isolated, medium-strength (EW $\sim$ 
10 to 200 m\AA) lines in the solar spectrum, using the solar atlas of \citet{WH98} as a guide. Atomic information for 
each line was retrieved from the VALD database \citep{KU99}. Log $gf$ values
given by VALD were then modified to force-fit abundances from ROBOSPECT analysis of solar spectrum EW to abundances
in the 2010 version of the abundance analysis software, MOOG \citep{SN73}. We used day-time sky spectra, taken as part of the
calibration data sets, as solar spectra for this normalization. MOOG analysis was conducted using the {\it abfind} driver and a solar model 
\citep{KU95} with the following atmospheric parameters:  ($T_{eff}, {\rm log\ } g, v_{t}, [Fe/H]$) = (5770 K, 4.40, 1.14 km/s, 0.00).

Our final line list contains 22 lines of interest (17 Fe, 3 Ni, 1 Ca, 1 Si), presented along with the relevant atomic
parameters in Table 2. To minimize the impact of blending, all selected lines are at least 0.5 \AA\ from any line with EW 
greater than 5 m\AA\ in the solar spectrum. To further minimize measurement distortions in spectra with lower SNR, the line
list input for ROBOSPECT included any significant feature within 2 \AA\ of a line of interest. EW for these features were
disregarded during the abundance analysis. The extraneous entries were added after observing unusual amounts of scatter among
the EW for lines with close neighbors, particularly at lower SNR. 

To evaluate the performance of our line list and to quantify the effects of a spectrum's SNR on EW values, we tested 
ROBOSPECT on 100 solar spectra, 25 each with mean SNR of 160, 130, 95 and 70.  These represent typical SNR for our program 
stars (see Fig. 2). We note that WH include an evaluation of ROBOSPECT's performance over a wide range of line strengths 
and SNR; our test results are consistent with theirs. The mean standard deviations in EW for our four test samples were 
4.2, 4.7, 7.5, and 11.1 m\AA, in order of decreasing mean SNR. The uncertainty this scatter introduces into our abundance 
results is small relative to the uncertainty due to external systematics. At all SNRs tested, 
we observed no significant correlation between the mean EW of a line and its associated uncertainty. Taken together, these 
results indicate that our line list is robust over the range in mean EW and SNR of interest.

The effect of potential errors in the determined radial velocity on EW measurement was evaluated by artificially increasing 
and decreasing our best estimate of radial velocity by a flat amount in the set of 100 solar spectra. For a shift of $\pm$2.5 km/s, 
corresponding to a radial-velocity uncertainty greater than that of 99\% of our program stars, the mean deviation in EW 
between the true and adjusted spectra was $-0.7$ m\AA.  When the shift was increased to 5.0 km/s, corresponding to a radial-velocity 
uncertainty greater than that of all program stars (including the rapid rotator WOCS 2016), the mean deviation in EW increased to $-3.5$ m\AA.  In both cases, 
ROBOSPECT correctly identified essentially the  same number of entries from the line list as it had with the correct 
radial velocity. We conclude that the effect of radial-velocity error on our abundance results is negligible.

Finally, we compared ROBOSPECT results with manual EW measurements for 18 random program  stars using a 10-line subset of our 
line list. Manually obtained EWs were measured using IRAF's  SPLOT tool. ROBOSPECT's results largely agree with the manual 
values, with ROBOSPECT's EW on average being 7.7 $\pm$4.2 m\AA\ (s.d.) lower than the manual results. We note that this consistent underestimate of EW by
ROBOSPECT (or, over-estimate using SPLOT and interactive measuring) is similar in solar spectra and in the spectra of program stars, and is not 
a function of SNR to any detectable extent, so we anticipate no effect on our abundance estimates based on differential analysis with respect to the sun. 

Using our measured radial velocities and constructed line list, EWs were measured for all 333 stars in our observing program. 
We discarded negative EW (ROBOSPECT's designation of emission lines) due to non-convergent fitting solutions, 
artifacts of cosmic ray removal, large noise spikes, or nonexistent lines in the measured spectrum. Approximately 4\% of 
the measured lines of interest returned negative values.

\section{Abundance Determinations}

Model atmospheres were constructed for each program star using the grid of \citet{KU95} and the input $T_{eff}$, 
log $g$ and microturbulent velocity values listed in Table 1.  As a reminder, each
star's $T_{eff}$ estimate is based on its $B-V$ color, individually dereddened based on its spatial position within
the field of the cluster.   For any star that is not actually a cluster member, this spatially interpolated reddening correction may be spurious; 
the gravity estimate based on an isochrone matched to the cluster CMD almost certainly would be as well. For suspected binaries, the color might reflect the combined light 
of the two stars and affect the assigned temperature.  While this might not be a large issue for SB1, a few candidate SB2 were highlighted by examination of spectra: WOCS 7009, 26007, 35025, 9004 and 60021.  As a conservative approach, we restricted our analysis to stars with membership probablity $\geq 50$\% (PL)
and no evidence of binarity.  We also excluded from our final analysis several stars with $(B-V)_0$ greater than 1.35, 
for which severe line-blending makes reliable measurement of equivalent widths problematic.   
This left us with a sample of 247 single, member stars.

Each star's measured equivalent widths and model serve as input to the {\it abfind} routine of MOOG to 
produce individual [A/H] estimates for each measured line for each star. A large 
volume of spectroscopic information is available for interpretation from over 7000 measured lines, with up to 17 [Fe/H] values for each star. 
Before constructing median [Fe/H] values for each star and for the sample as a whole, the following filters were applied to the individual abundance estimates: 
abundances based on equivalent width measurements smaller than 
three times the expected error in equivalent width for each star based on the
star's SNR, were not considered, nor were equivalent widths
for very strong lines (EW $\geq 200$ m\AA). We note that the consequences of using 200 or 250 m\AA\ as the upper limit for EW are negligible. 
Estimated abundances from individual lines deviant by more than 1 dex from solar were not included 
in the determination of [Fe/H] for each star. 

Figures 2 and 3 show the spread of stellar [Fe/H] estimates for each star as a function of the 
individual spectrum SNR and unreddened $B-V$ color for each star. A few outliers with SNR below 
50 are obvious, as are a few of the brightest (and coolest) stars with higher than typical abundance.
Figure 3 suggests that abundances for the cooler stars appear to be slightly below 
the range for the stars nearer the turnoff. We further divided our sample of 
247 likely members into stars bluer than and redder than $(B-V)_0 = 0.60$ to separate out the cooler 
and more evolved stars and re-examined the effect of filters that precede the construction of
each star's median [Fe/H] abundance.  Do some of our 17 iron lines produce abundances in dwarfs that are unacceptably noisy?  

By examining the effect of imposing different criteria for excluding small EW measures, we were able to highlight several iron lines, which for dwarfs, lead to biased [Fe/H] values in the sense that the abundance values for these lines in dwarf stars are higher than the ensemble median value by several tenths. The reason for this is clear;  if the median EW for a particular absorption line is small, imposing a 3-sigma cut will preferentially preserve stars in the sample with larger EW, and consequently larger [Fe/H] values.  We excluded lines in dwarfs for which imposing 3-sigma cuts changed the median EW for the line by more than 25\% in comparison with a more generous 1-sigma cut.  As a reminder, these lines are measured in the solar spectrum at considerably higher SNR so we may still place considerable confidence in the log $gf$ values determined from those lines.   

As this still leaves information from several thousand separate iron line measurements, we were able to further explore the consequences of these filters.  
A diagnostic diagram is presented in Figure 4, demonstrating the lack of a trend in the median 
[Fe/H] for each of our 17 iron lines with wavelength, where the median value is constructed from the full sample of 247 probable-member, single stars for lines denoted by
black symbols; the six lines denoted by red symbols represent measures for giant stars only. 
The upper panel of this figure shows the MAD ({\it median absolute deviation}) statistic for each line, a robust estimator of variance, constructed by comparing the 
[Fe/H] value for each star to the median value for that particular absorption line.  

For the subset of 247 single member stars bluer than $(B-V)_0 < 1.35$, the median [Fe/H] value is 
$-0.03 \pm 0.06$, with abundances for the other elements that are entirely consistent with solar: 
$-0.01 \pm 0.10, -0.01 \pm 0.06 $ and $0.00 \pm 0.06$ for [Ca/H], [Si/H] and [Ni/H], respectively.  Of these 247 stars,
200 are main sequence and turnoff stars, with a median iron abundance of $-0.02 \pm 0.05$,
 while the remaining 47 cool and evolved stars show a lower [Fe/H], $-0.09 \pm 0.05$.  
We note that this apparent discrepancy between giant and dwarf [Fe/H] values is nearly identical to that found
in NGC 3680 \citep{AT09}.  
  
For all of these quantities, the indicated error is the MAD statistic; for a large sample size 
drawn from a normally distributed population, the normalized MAD statistic, MADN = 1.48 MAD, 
approximates well the sample standard deviation, permitting an estimate of the traditional standard error of the 
mean through division by $(N-1)^{1/2}$. For the sample of 247 single, probable-member stars, the MAD statistic implies a value for MADN = 0.09; as this
statistic is a good proxy for a standard deviation, a standard error of the mean of 0.01 is implied.  These values can be confirmed by computing averages, standard deviations and s.e.m. values in linear space; for these stars, the average [Fe/H], computed in linear space and then converted to logarithmic values, is $-0.02^{+0.07}_{-0.13}$ (standard deviations).   

Table 3 includes the median [Fe/H], along with the number of iron lines included in the estimate, as well as estimates of [Ca/H], 
[Si/H] and [Ni/H] for 333 stars. Stars not part of the set of 247 single, probable members stars are flagged with a note indicating non-member of binary status.
The statistic that accompanies each [Fe/H] estimate is the MAD. 
We note that the 29 stars not included in these estimates of cluster [Fe/H] due to evidence of binarity, yield
a similar median [Fe/H] value if the suspected double-lined spectroscopic binaries are excluded.  For member stars which are mostly SB1, 
[Fe/H]$ = -0.05 \pm 0.06$, consistent with the more conservative sample that excludes binaries.  
  
Conversions of photometric parameters to $T_{eff}$ and $V_t$ are fundamentally discontinuous for dwarf 
and giant stars; to assess the effects of parameter choices on the derived 
abundances, each star was reanalyzed with atmospheric parameters incremented and decremented by 
the following amounts: 100 K for $T_{eff}$, 0.25 for log $g$ and 0.25 km/sec for $V_t$.  A summary 
of the extensive results is shown in Figure 5 where once again median values and variances 
indicated by median absolute deviations are compiled and shown for stars in different color bins. 
This figure illustrates some well-known dependences of abundance determinations on stellar parameters, 
namely that main sequence star abundances are very sensitive to $T_{eff}$ estimates, while evolved 
stars demonstrate a higher sensitivity to surface gravity and consequently to microturbulent velocity 
estimation schemes.  

Although the discrepancy between the giant and near-turnoff samples is small, it is of value to 
explore modest parameter changes that could eliminate it.  Reference to Figure 5 suggests that errors 
in log $g$ have relatively little impact; we remind the reader that the log $g$ values employed are 
consistent with values suggested by asteroseismic analyses of NGC 6819 giants. The relation used for 
microturbulent velocities for the evolved stars is more tenuous; a fairly modest decrement of 0.2 km/sec 
would raise the giant abundances to match the [Fe/H] representative of the near-turnoff stars.  
Finally, increased abundances for both regions of the CMD would accompany higher temperatures, though a 
simple increment to the mean cluster reddening will have differing effects for the two classes 
of stars, both because of the slight color-dependence of the effect on the abundance as shown 
in Figure 5 and the larger increment implied for warmer stars resulting from the same higher reddening.   

To define the final estimate of [Fe/H] for the cluster, we default to the dwarfs, where the sensitivity
to the changes in the three key input parameters ($T_{eff}$, $V_t$, log $g$) is collectively minimized,  
the color range is small, and the parametric differences between the program stars and the solar
reference spectra are significantly smaller than for the evolved stars. For log $g$ and $V_t$ we 
adopt 0.1 dex and 0.2 km/sec as generous estimates
for the uncertainties. From Fig. 5, these translate into 0.003 and 0.008 dex, respectively. As usual, the
more sensitive parameter is the temperature, tied to the $B-V$ color transformation, which has two
primary components, the uncertainty in the mean cluster reddening, $E(B-V)$, and the uncertainty in the
photometric zero point. From Paper I, the photometric internal and external errors combined lead to an
uncertainty of 0.007 in $E(B-V)$, while the photometric merger of the four broad-band surveys above
implies an uncertainty in the $B-V$ zero-point of 0.004 mag. Combined, the $(B-V)_0$ scale is reliable
to 0.008 mag. On our temperature scale, this translates to approximately 32 K or, from Fig. 5, 0.012 dex
in [Fe/H]. Combining all three parameters, one arrives at a final external uncertainty of 0.015 dex, 
much larger than the internal precision of the spectroscopic average. From 200 turnoff stars, the mean
[Fe/H] for NGC 6819 is determined to be $-0.02 \pm 0.02$. 

Among the prior determinations of NGC 6819's heavy element abundance, the result published by \citet{BR01} 
has anchored a string of moderately super-solar values. Their study of three clump 
giants utilized the high dispersion ($R \sim 40,000$) spectrograph on the Galileo Italian 
National Telescope (TNG). Though the sample size was small, their analysis incorporated hundreds of lines, 
including nearly 100 iron lines, six of which were Fe II lines to permit an accurate verification of 
surface gravity.  The reddening value of E$(B-V) = 0.14$ they adopted, however, was somewhat smaller than the 
mean value derived in Paper I and utilized here, E$(B-V) = 0.16$.  We repeated parameter determinations 
for the three giants studied by \citet{BR01}, including independent determinations of spatially-dependent 
reddening and $T_{eff}$ values, log $g$ values from isochrone comparisons and microturbulent velocity 
values based on the surface gravities. Using the estimates of $\Delta$[Fe/H] for each of the principal 
atmospheric parameters cited by \citet{BR01}, their equivalent width analysis would yield 
an average abundance 0.07 dex lower with our atmospheric parameters, lowering their [Fe/H] for the cluster 
from $+0.09$ to $+0.02 \pm 0.03$, essentially solar.

\section{Discussion and Conclusions}
High dispersion spectra of 333 potential members of the old open cluster, NGC 6819, have been processed and analyzed. From
the wavelength region near Li 6708 \AA, abundances have been obtained for Fe, Ni, Ca, and Si for 247 highly probable, single-star members
ranging from the tip of the giant branch to well below the top of the CMD turnoff. Using up to 17 Fe lines per star, the 
cluster exhibits a slightly subsolar [Fe/H], a conclusion which remains unchanged if the data are sorted into evolved stars (47) or
turnoff/unevolved stars (200). Due to the cumulative impact of a few thousand Fe lines, the standard error of the mean for [Fe/H] 
tied to internal errors is below 0.01 dex, leaving the systematics of the temperature, log $g$, and microturbulent velocity scales 
as the overwhelming source of uncertainty in the final cluster metallicity. Once again, thanks to the size of the sample, we can 
minimize the impact of these parameters by looking only at the 200 unevolved stars since the red giant abundances have a higher 
degree of sensitivity to microturbulent velocity. From these stars alone, [Fe/H] = $-0.02 \pm 0.02$, where the uncertainty includes 
both internal and external errors. The lower metallicity compared to the commonly adopted value of [Fe/H] = +0.09 \citep{BR01} continues 
a trend first indicated by the intermediate-band photometry of Paper I, where precision photometry was used to derive a reliable cluster 
reddening and confirm the existence of variable reddening across the face of the cluster (PL), leading to subsolar [Fe/H] from either
$m_1$ or $hk$ indices.

Within the context of Galactic evolution, NGC 6819 exhibits no features which distinguish it from a typical thin-disk population 
formed within the last 5 Gyrs within 1 kpc of the Sun's galactocentric distance. Its [Fe/H] places it at the mean for open
clusters at the solar galactocentric radius, a distribution which exhibits a scatter at only the $\pm$0.1 dex level \citep{TA97}.
Its abundance ratios relative to Fe are all consistent with solar, a pattern which holds true within the scatter for Ni, Ca, and 
Si for stars with [Fe/H] $\sim$0.0, whether they are classed as members of the thin or thick disk population (see \citet{HI14} 
and references therein). NGC 6819 forms a rich and ideal link for tests of stellar evolution theory midway in age between 
NGC 5822 \citep{CA11} and M67 \citep{SA04}, a characteristic which will prove valuable in the analysis of the evolution of Li
for intermediate mass stars evolving off the main sequence and up the giant branch.
 
\acknowledgments
The authors express their thanks to Imants Platais for supplying additional data to compute spatially-dependent reddening estimates in the cluster. 
Drs. Ryan Maderak and Jeff Cummings assisted as visiting astronomers at KPNO for runs in 2010, supplying support and advice for the spectroscopic data collection. NSF support for this project was provided to ER while he was an undergraduate at the University of Kansas 
through NSF grant AST-0850564 via the CSUURE REU program at San Diego State University, 
supervised by Eric Sandquist.  NSF support for this project to BJAT, BAT and DLB through NSF grant AST-1211621, and to CPD through NSF grant AST-1211699 is gratefully acknowledged. Extensive use was made of the WEBDA \footnote{http://www.univie.ac.at/webda} database operated at the Department of Theoretical Physics and Astrophysics of the Masaryk University,
the TOPCAT suite \footnote{http://www.star.bristol.ac.uk/~mbt/topcat/} the Vienna VALD, of the MOOG suite of spectroscopic analysis software, 
and of ROBOSPECT software.

\clearpage
\begin{figure}
\makebox[\textwidth]{
\includegraphics[angle=270,scale=0.6]{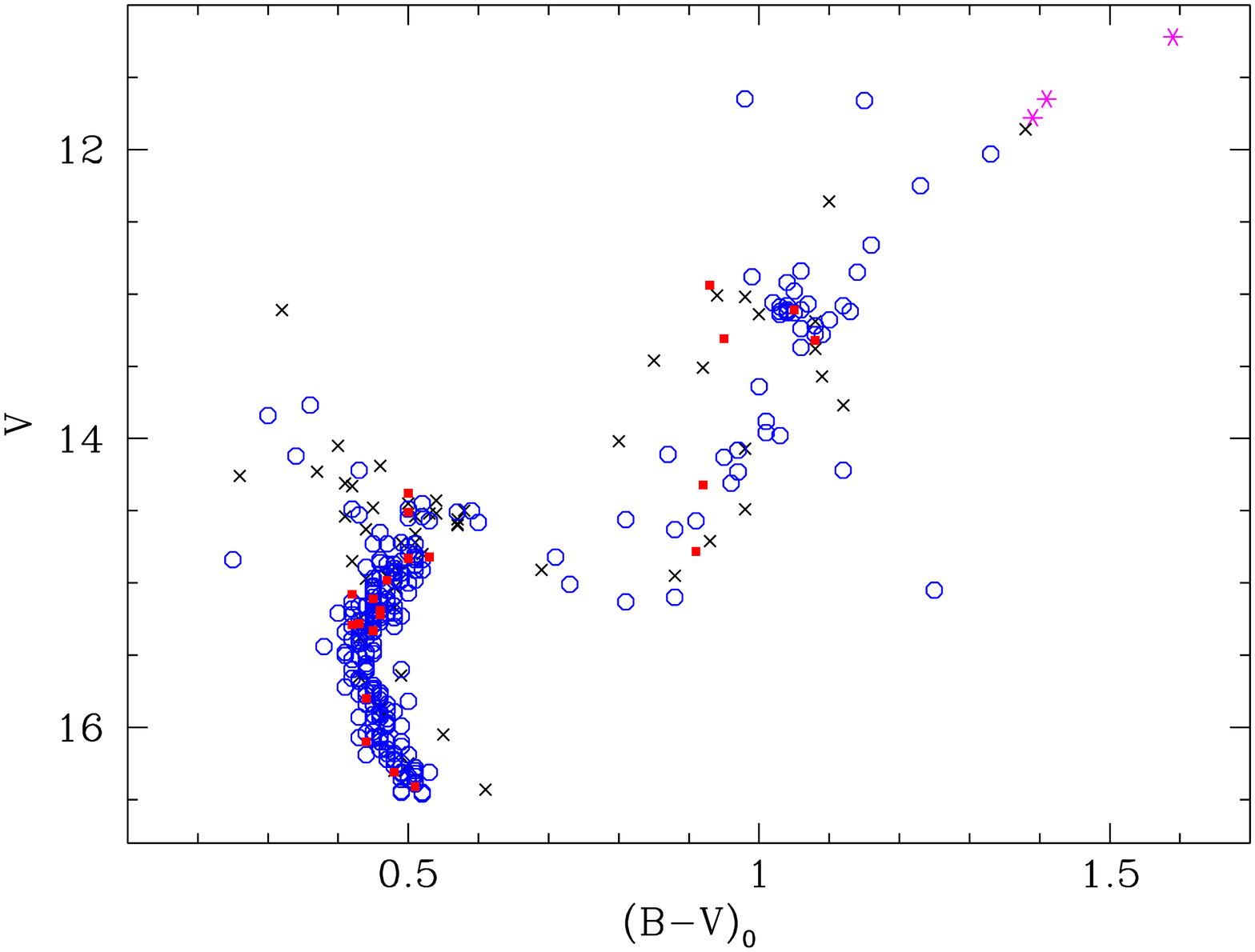}}
\makebox[\textwidth]{\caption{Color-magnitude diagram of spectroscopic sample.  Blue open circles denote single, probable member stars.  The few large magenta asterisks mark stars with very cool colors, while black crosses mark probable nonmembers and red filled squares denote probable member spectroscopic binaries.}}
\end{figure}
\clearpage
\begin{figure}
\makebox[\textwidth]{\includegraphics[angle=270,scale=0.6]{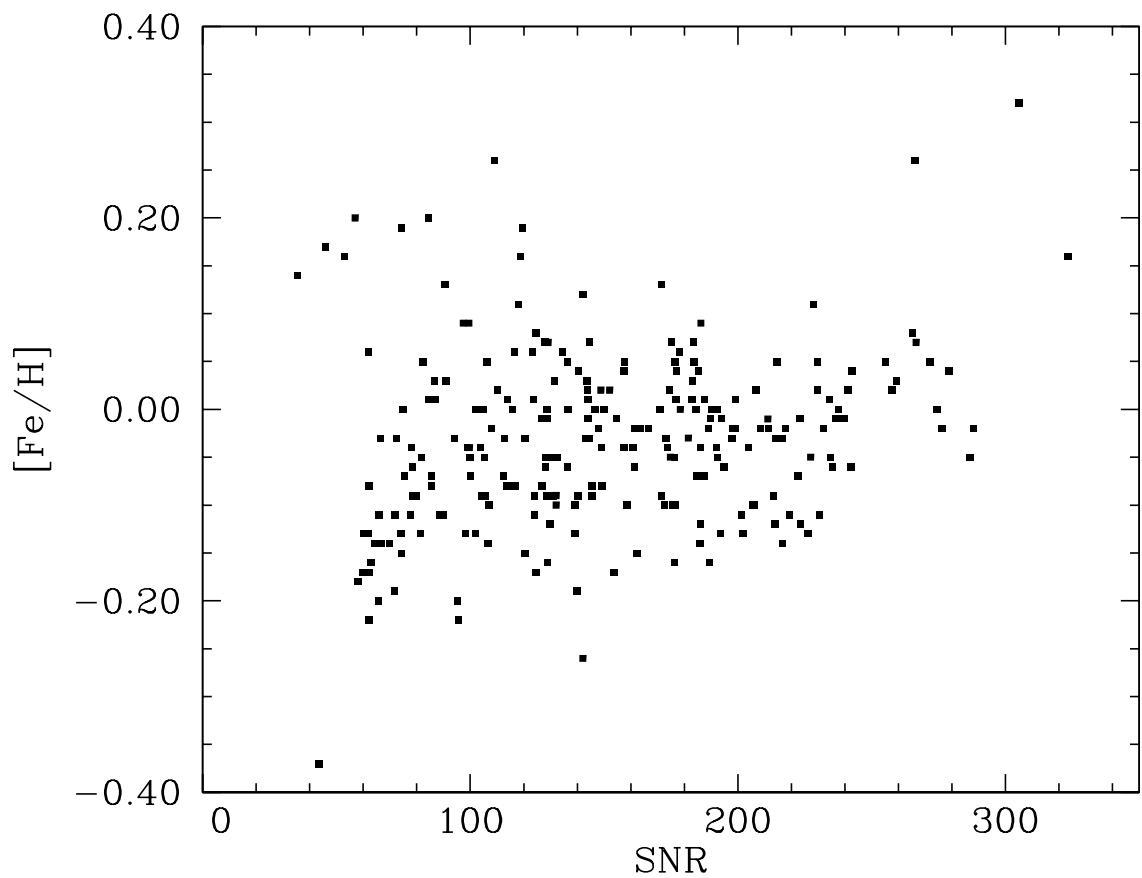}}
\makebox[\textwidth]{\caption{For 247 single, probable member stars in NGC 6819, the median iron abundance relative to solar is shown as a function of the spectrum SNR.}}
\end{figure}
\clearpage
\begin{figure}
\makebox[\textwidth]{\includegraphics[angle=270, scale=0.6]{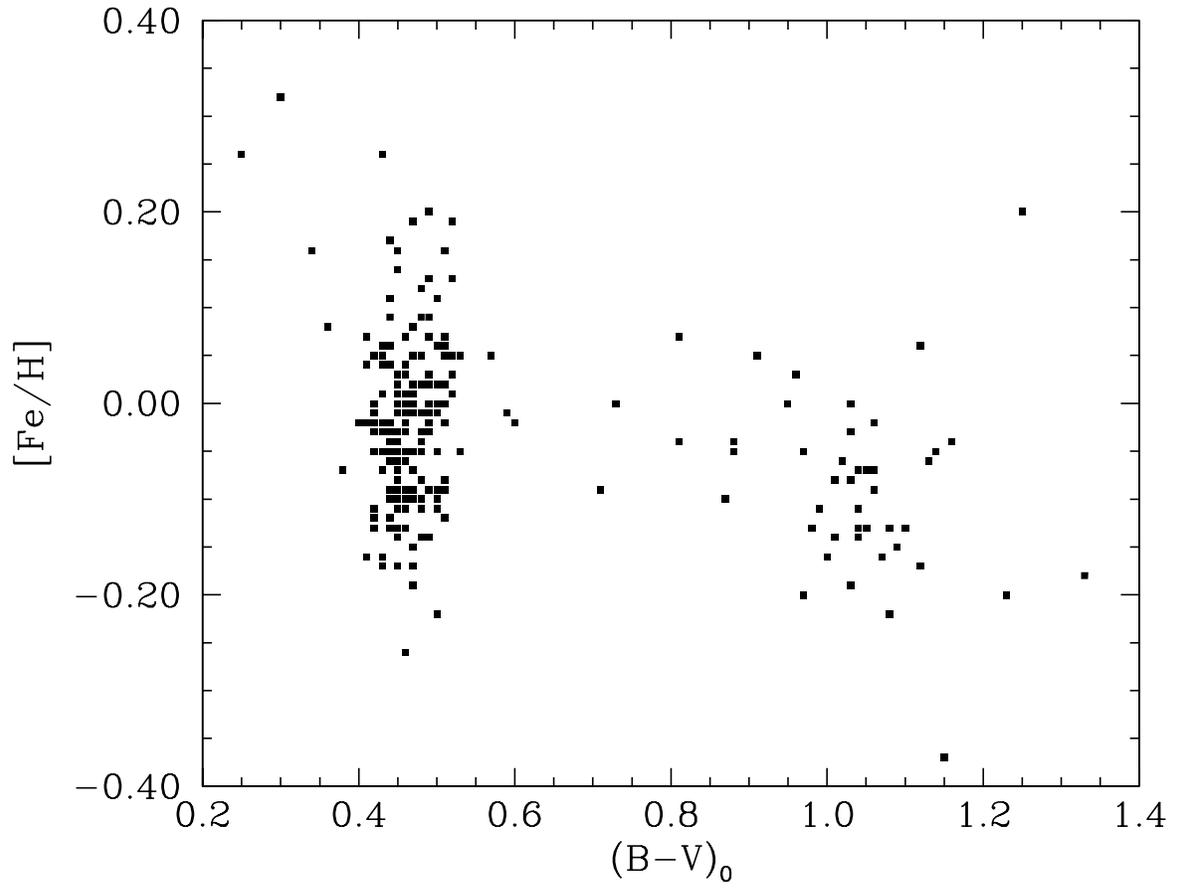}}
\makebox[\textwidth]{\caption{For 247 single, probable member stars in NGC 6819, median iron abundances relative to solar is illustrated as a function of reddening-corrected color, $(B-V)_0$. }}
\end{figure}
\clearpage
\begin{figure}
\makebox[\textwidth]{\includegraphics[angle=270,scale=0.6]{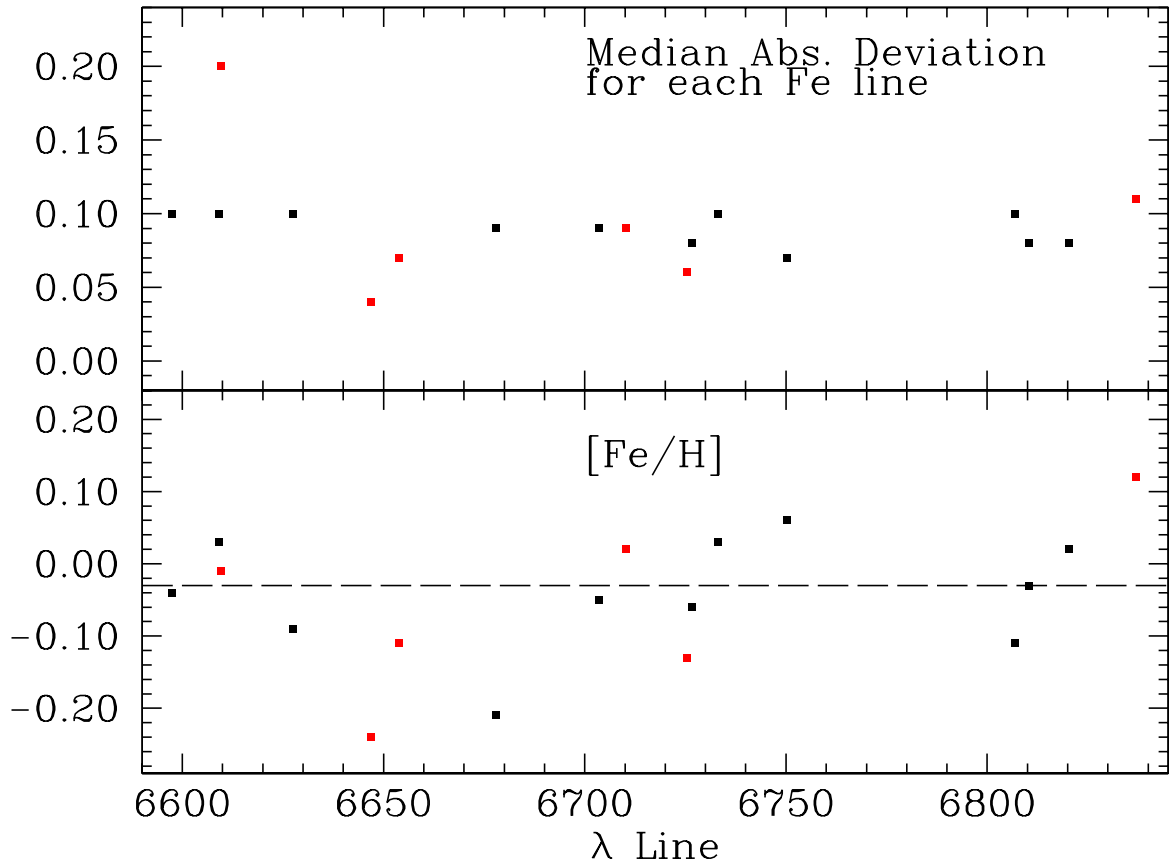}}
\makebox[\textwidth]{\caption{For single, probable member stars, the median abundance is shown for each of the 17 iron lines used, plotted as a function of wavelength expressed in \AA.  The upper panel shows the median absolute deviation of values about each line's median value. Values denoted by red symbols indicate lines for which only giant stars contribute to the median abundance value.
In the bottom panel, the dashed horizontal line denotes the cluster [Fe/H] value of -0.03.}}
\end{figure}
\clearpage
\begin{figure}
\makebox[\textwidth]{\includegraphics[angle=270, scale=0.6]{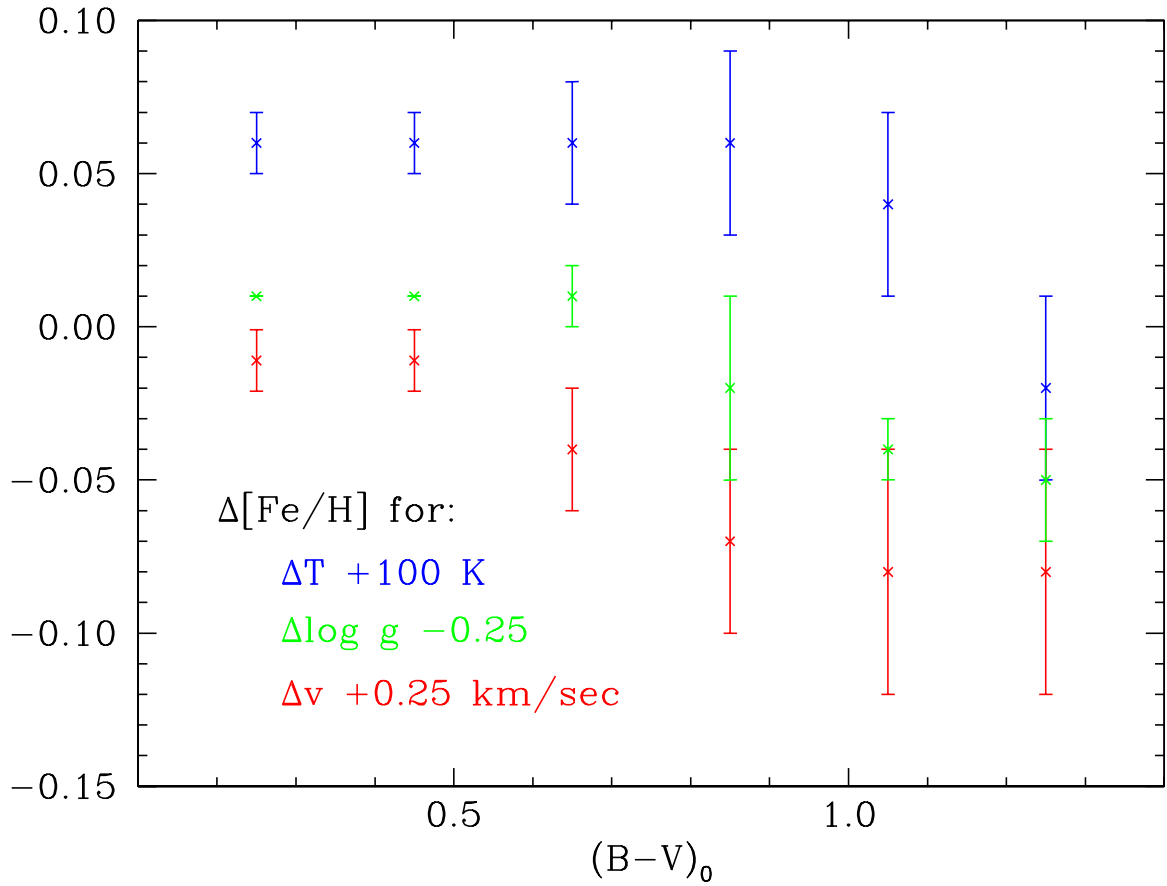}}
\makebox[\textwidth]{\caption{Median values of $\Delta$[Fe/H] for stars in unreddened color bins 0.2 mag wide illustrate the effects of increasing the model atmosphere temperature by 100 K (blue filled squares), of increasing the microturbulent velocity parameter by 0.25 km/sec (red cross symbols) and lowering the surface gravity parameter, log $g$ by 0.25 (green open triangles). }}
\end{figure}

\clearpage
\begin{figure}
\makebox[\textwidth]{\includegraphics[angle=270, scale=0.9]{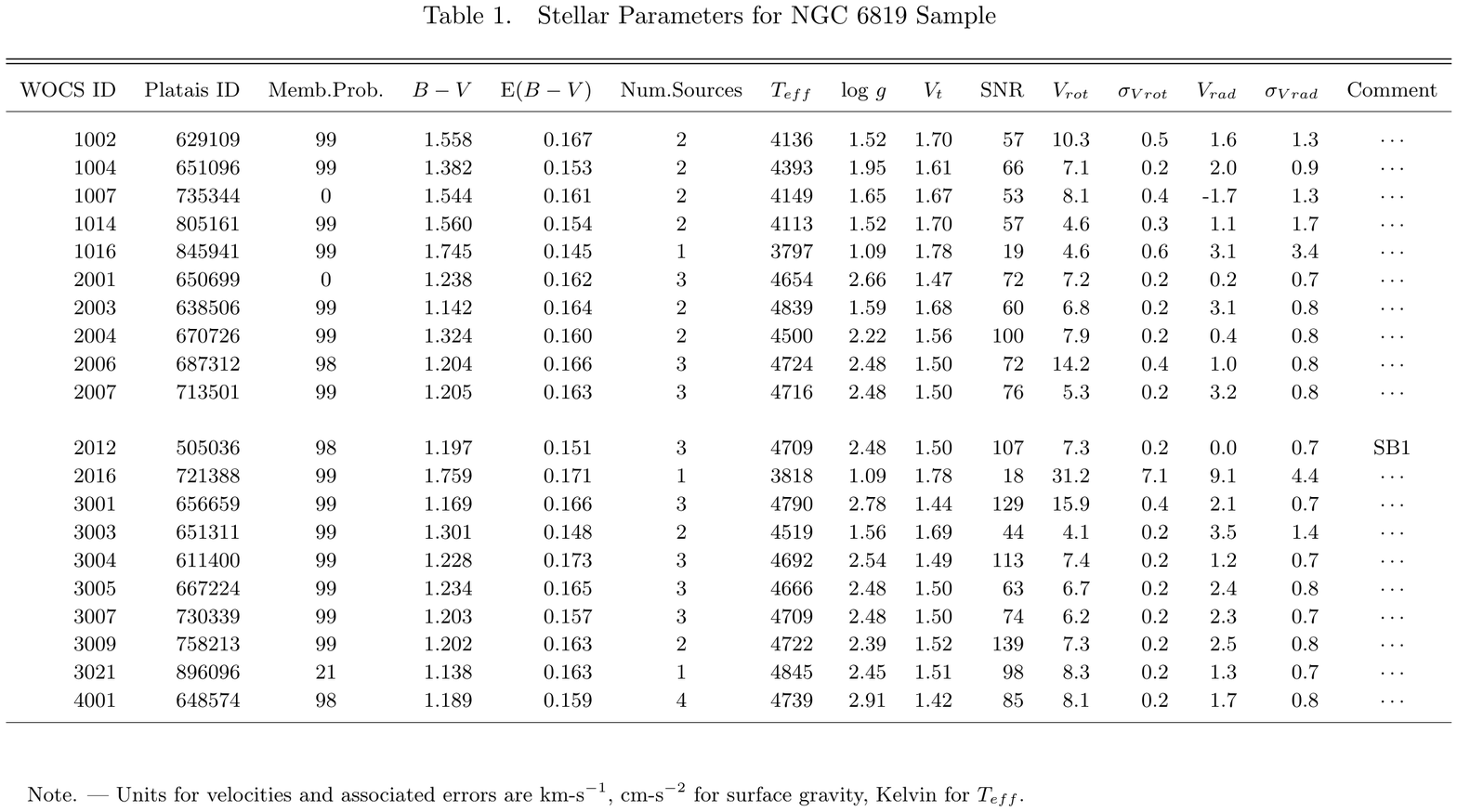}}
\end{figure}
\clearpage
\begin{figure}
\makebox[\textwidth]{
\includegraphics{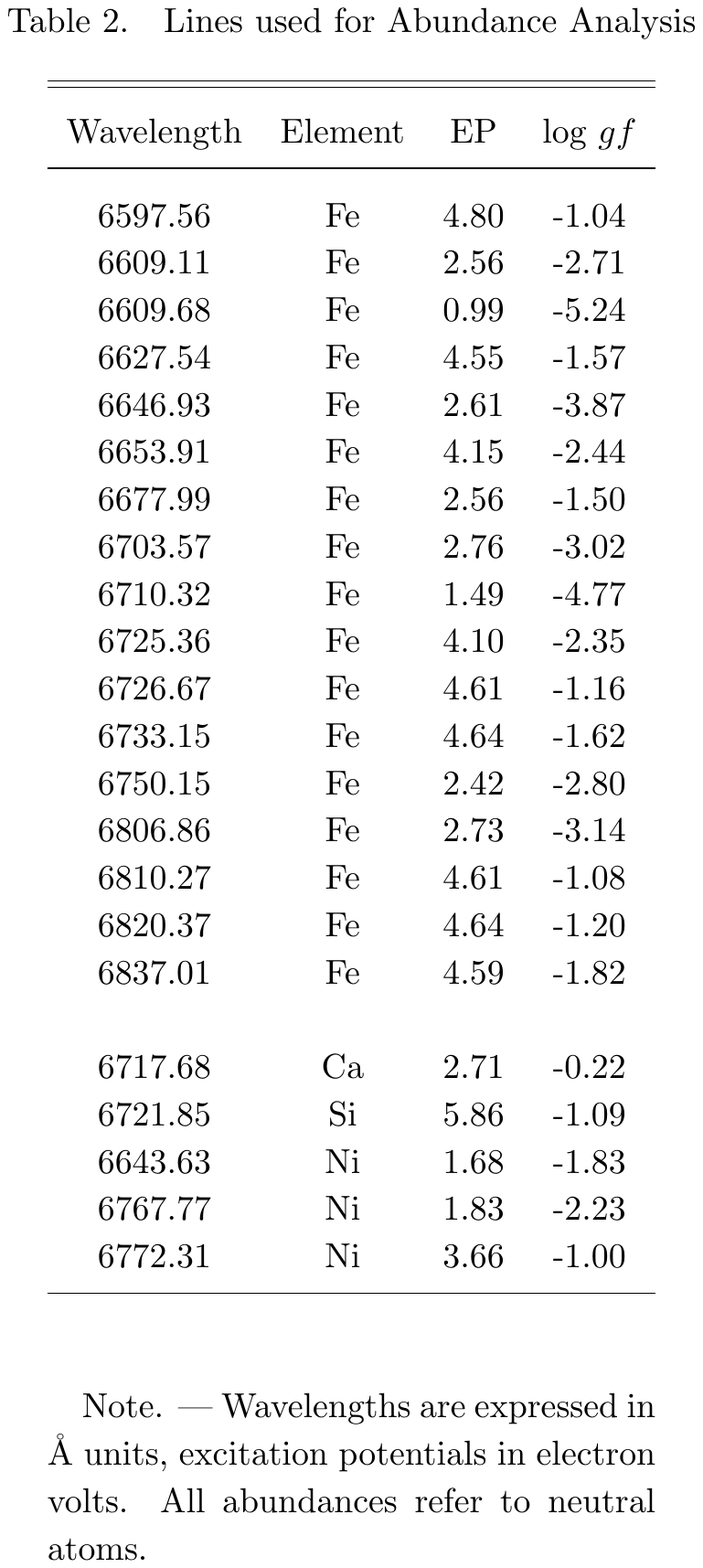}}
\end{figure}
\clearpage
\begin{figure}
\makebox[\textwidth]{\includegraphics{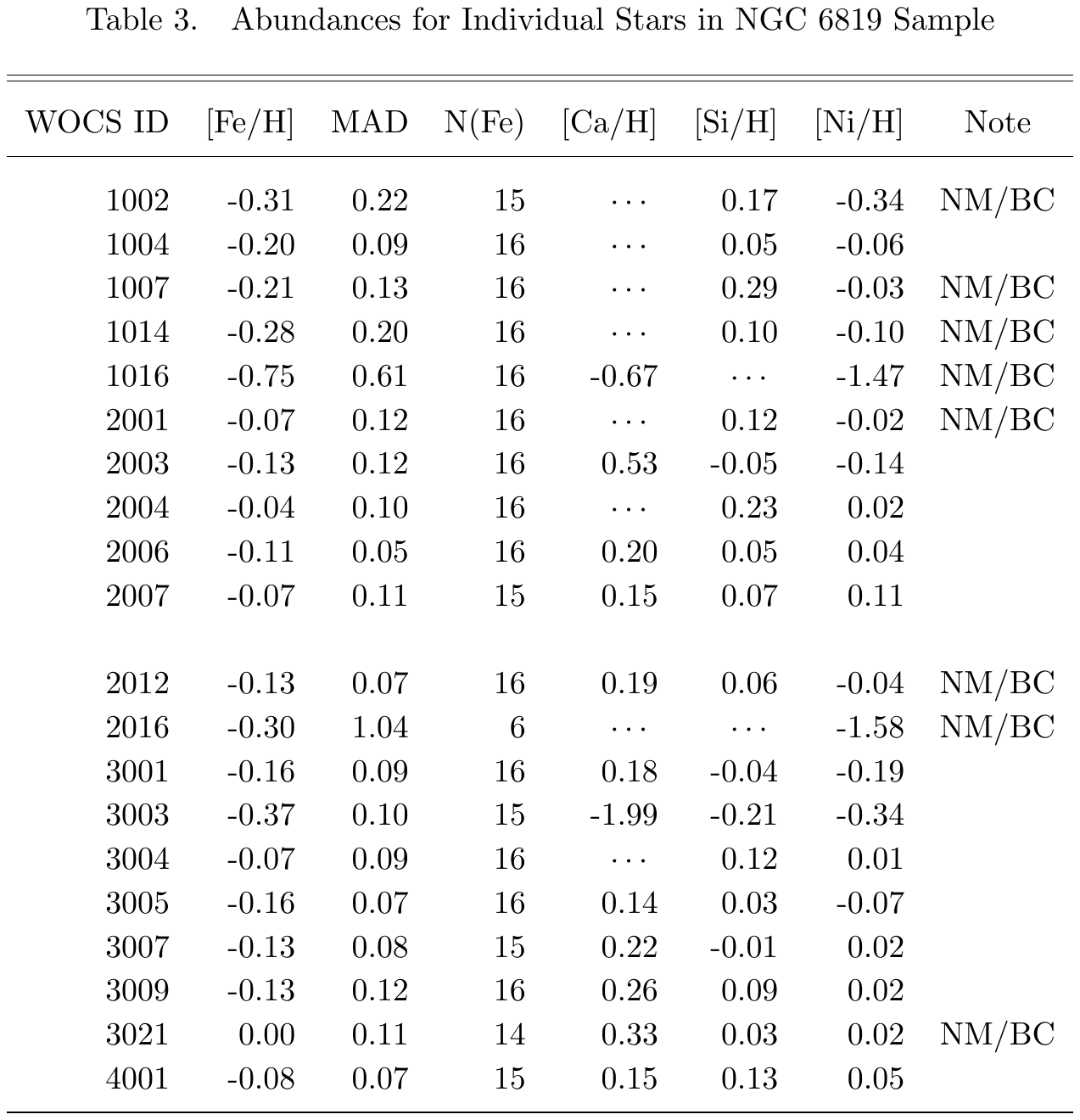}}
\end{figure}
\end{document}